# Phase stability, local chemical disorder and its effect on the mechanical properties of refractory high-entropy alloys

Soumyadipta Maiti [a, *], Walter Steurer [a]

[a] *Laboratory of Crystallography, Department of Materials, ETH Zurich, Vladimir-Prelog-Weg 5, CH-8093 Zurich, Switzerland*

## Abstract

Equiatomic TaNbHfZr and TaNbHfZrTi high-entropy alloys (HEAs) have been investigated for their phase stability, short-range clustering (SRC) type chemical ordering, other structural phenomena and and the effects on strengthening. Both the HEAs show body centered cubic (bcc) average structures in their as-cast and annealed state (1800°C for 1 day). Established thermodynamic and atomic size related parameters governing the general phase stability of HEAs are applied to investigate into the single-phase formability. *Ab-initio* calculations have been done to determine the thermodynamic stability parameters. The as-cast structures do not contain any local chemical order, but the high-resolution transmission electron microscopy (HRTEM) studies reveal their locally distorted lattice. The local lattice distortion of the average structure in their as-cast and annealed state is quantified by refining atomic displacement parameters (ADPs) from single-crystal X-ray diffraction (XRD) and powder neutron diffraction (ND) studies. The static component of the ADPs appears to dominate over the thermal ADP, suggesting a high degree of local lattice distortion in the HEAs. The lattice distortions cause solid-solution like strengthening, which increases the hardness of the HEAs by a factor of 2-3 compared to the rule of mixture. The SRCs that form in the annealed HEAs have a planar profile perpendicular to the <1 0 0> directions. The SRCs mainly have a Gaussian-like in-depth concentration profile across their width with clustering of Zr and Hf. The SRCs have a higher atomic volume causing local tetragonal lattice distortions, which in terms produce diffuse streak-like intensities in the XRD datasets and in the fast Fourier transform (FFT) of the HRTEM lattice images. The presence of SRCs locally reduces the lattice potential energy of the structure and puts additional energy barrier to the movement of dislocations. This increases the hardness causing around 50% of additional strengthening in the annealed HEAs. Thermodynamic analysis shows that the relaxed structure containing the SRCs lower the enthalpy and the configurational entropy. The change in free energy still remains negative and allows for the presence of SRCs.

[*]Corresponding author: S. Maiti, Tel: +41 44 633 71 29, E-mail: smaiti@mat.ethz.ch

# 1. Introduction

High-entropy alloys (HEAs) are single-phase solid-solution alloys of multiple principal elements, usually with 4 or more elements [1, 2]. HEAs suppress the formation of possible intermetallic phases due to the large configurational entropy and are also thought to have a distorted local lattice structure [3, 4]. This in terms is believed to be responsible for some of the very interesting properties of HEAs like their use in high-strength structural materials for high-temperature and ambient-temperature applications [5, 6], high fracture toughness at ambient and cryogenic temperatures [7-9], wear resistance [2, 10], sluggish diffusion etc [11]. Due to all its interesting properties, the HEAs have an increasing number of publications over the recent years [12]. Single-phase HEAs usually form a body-centered or face-centered cubic (bcc or fcc) structure and the structural stability of the as-cast and homogenized alloys are found to be dependent on a thermodynamic and an atomic size variation based parameter [13]. In the literature the thermodynamic parameter involved in the calculation of enthalpy of mixing is often calculated by the method applicable for metallic glasses and the approach may not be applicable for the crystalline distorted matrix of the HEAs [13-15]. The local short-range ordering or clustering (SRO/ SRC) present in the solid-solution structure can change the enthalpy [16] and the configurational entropy [17] of the system affecting the Gibbs free energy.

In recent years refractory based HEAs have gained much importance for their potential application in the high demand area of materials for high-temperature aerospace applications [18, 19], ambient-temperature high-strength structural materials [20] and materials for micro and nano-electromechanical (MEMS, NEMS) devices [6]. For the long-term high-temperature applications of HEAs, it becomes very important to understand the nature of the local chemical ordering phenomena of the single-phase average structure. To the authors' knowledge, until now the nature of long-term evolution of the SRC has been only studied for the TaNbHfZr system and there exists a general deficiency of knowledge in the long-term phase stability, thermodynamic parameters, local structural and chemical ordering and their effects on the energetics and the properties of the HEAs [21].

For this detailed study of the nature of chemical ordering, equiatomic TaNbHfZr and TaNbHfZrTi HEAs were chosen because it produced detectable weak diffuse scattering in X-ray and electron diffraction as an indicative of the presence of local chemical ordering in crystals [22]. The HEAs have been studied for their stability criteria for the formation of single-phase average structure and the local lattice distortions in the as-cast structure. Further, the detailed investigation of the chemical ordering in annealed materials, local structural relaxations and effects on microhardness have been discussed.

## 2. Experiments

The equiatomic TaNbHfZr and TaNbHfZrTi HEAs were prepared by arc-melting together the different constituting elements. Ta and Nb were taken in powder form (99.98 % and 99.99% producer) and pre-alloyed together. Later the other elements, Hf, Zr and Ti were alloyed from high-purity solid pieces (> 99.8 %, oxygen level ~ 140, 130 and 1300 ppm, respectively). The casts of 5-6 grams were re-molten 5 times, each time flipped upside down and a Ti getter was used to consume traces of oxygen inside the arc furnace chamber. The HEAs were annealed at 1800°C up to a maximum of 1 day inside sealed Ar filled Ta ampoules. The annealing temperature for TaNbHfZr and TaNbHfZrTi alloys were 0.78 and 0.82 times the expected melting temperature from the rule of mixture, which is generally observed and used in the HEAs [13, 18]. An equiatomic Hf-Zr alloy was put inside the Ta ampoules to be used as additional oxygen getter during homogenization and annealing treatments. To indicate the two different HEAs in this work, TaNbHfZr and TaNbHfZrTi alloys are termed as HEA-4E and HEA-5E, respectively.

The average bcc structure of the alloys was first identified by using an in-house powder diffractometer (PANalytical X'Pert PRO diffraction system) with Cu $K_{\alpha 1}$ monochromatic radiation in a 2θ (diffraction angle) range from 20° to 120°. Single-crystal X-ray diffraction (XRDs) datasets were collected by an in-house diffractometer (Oxford Diffraction, Xcalibur system with Mo $K_\alpha$ radiation) to quantify the atomic displacement parameters (ADPs). Synchrotron radiation (0.6836 Å wavelength with PILATUS 2M detector at SNBL, ESRF, Grenoble) was used to investigate into diffuse scattering and local structural oredring. Powder neutron diffraction (ND) was done on the homogenized HEAs using a wavelength of 1.1545 Å (HRPT diffractometer, Paul Scherrer Institute, PSI, Villigen, Switzerland). This provided additional advantage to quantify the ADPs and to detect any long-range order present in the structure by the differences of neutron scattering lengths (scattering lengths for Ta, Nb, Hf, Zr and Ti are 6.91, 7.05, 10.9, 7.16 and -3.44 fm, respectively). Further details of the diffraction experiments can be found elsewhere [21].

The microstructure and bulk compositions were analyzed by a scanning electron microscope (SEM, SU-70 Hitachi) in backscattered electron (BSE) mode, combined with an energy-dispersive X- ray spectroscope (EDX, X-MAX Oxford Instruments). The local structures of the HEAs were investigated by high-resolution transmission electron microscopy (HRTEM). An FEI Tecnai F30 and an FEI Talos instrument equipped with field emission guns were used for the HRTEM operating under 300 and 200 kV, respectively. The details of the sample preparation for SEM-EDX and the HRTEM investigations are published elsewhere [6].

The local chemical compositions at the SRCs were investigated by the use of atom probe tomography (APT) techniques. A LEAP 4000X-HR system equipped with the Reflectron geometry for enhanced mass-resolving power of was used for the APT runs. The probes from the HEAs for APT were fabricated by focused ion-beam (FIB) of Ga ions following a standard 'lift-out procedure' by using a FIB-SEM system (Helios Nanolab 600i) [23]. The prepared APT tips mounted on the top of Si posts had a tip radius of less than 50 nm and were subjected to run in the APT chamber in laser pulsing mode. The temperature of the probes were kept in the 33-64 K range and operated at a local electric field of around 45 V/nm. The laser pulse had an energy of 100 pJ and pulsing frequency of 160 kHz.

Vickers microhardness were measured on the HEA surfaces by applying a 200 g load over at least 15 different places. The microhardness were taken for HEAs in their as-cast state and also for states annealed for different time-periods, a maximum up to 1 day of annealing.

## 3. Results and discussion

### 3.1 As-cast state, phase stability and local lattice distortions

The two HEAs in their as-cast state show a bcc average structure with a dendritic microstructure as revealed by powder XRD and microstructures published in the literatures [21, 24]. In this work, the HEAs have been analyzed for their propensity to form single-phase solid-solutions, based on the parameters that have been found to govern for the stabilization of single-phase structures [13]. A thermodynamic and an atomic size variation based parameter, $\Omega$ and $\delta$ have been derived in the literature from the qualitative analysis of the stability criteria of the HEAs as given by

$$\Omega = \frac{T_m \Delta S_{mix}}{|\Delta H_{mix}|}$$

where $\Omega$ is the ratio of entropic contribution to the Gibbs free energy with respect to the enthalpy of mixing, $\Delta H_{mix}$, $\Delta S_{mix}$ is the configurational entropy of mixing, $T_m$ is the expected melting temperature or the annealing temperature of the HEA and

$$\delta = \sqrt{\Sigma_{i=1}^{n} c_i \left(1 - \frac{r_i}{r}\right)^2}$$

where $\delta$ is the parameter for the variation of atomic radii of the elements, $r_i$ is the atomic radius of the component $i$, $r$ is the average radius of elements and $c_i$ is the concentration of element $i$. The values of $\Delta H_{mix}$ for the alloys were calculated from ab-initio based methods and described in details in section 4 of thermodynamic modeling. Based on the thermodynamic and atomic-size variation parameters, it appears that the two HEAs should be well inside the solid-solution forming region in the phase stability map, defined by the conditions: $\Omega \geq 1.1$ and $\delta \leq 6.6\%$. This is shown in Fig. 1 of the stability regions of different kinds of multi-component materials like HEAs, intermetallics and bulk metallic glasses (BMGs) compiled together. The details of these parameters for the two HEAs are given in Table 2 in section 4. The valence electron concentrations (VEC) of HEA-4E and HEA-5E are 4.5 and 4.4, respectively, which is also in accordance with the observed statistics of HEAs with VEC ≤ 6.87 tending to form bcc solid solutions [12, 25]. The two studied HEAs are also expected to remain stable bcc phases after the homogenization and annealing treatment, because the high annealing temperature (1800 °C) would still keep the thermodynamic parameter, $\Omega$ high in the solid-solution region.

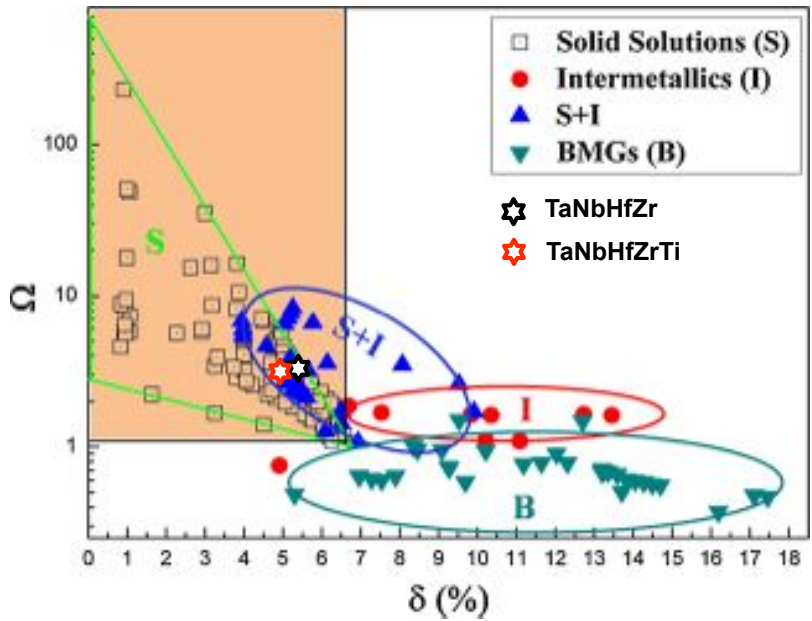

Fig. 1. Position of the TaNbHfZr and TaNbHfZrTi alloys with respect to the of solid-solution formation regions based on the thermodynamic and atomic-size based parameters, $\Omega$ and $\delta$. Figure produced with permission from Elsevier [13].

As the studied HEAs formed a dendritic structure in the as-cast state, it was not possible to obtain a homogenous single-crystal with fixed orientation to be studied for the local disorder by XRD and diffuse scattering [22]. So, to complete the study of the nature of the local structure of the HEAs, HRTEM investigations and selected area electron diffractions (SAED) were done for the as-cast HEA-4E. A study of the HEA-5E is already available in the literature [24]. Fig. 2 shows the as-cast local structure of HEA-4E by HRTEM in bright field (BF) mode. The SAED pattern of the local structure oriented along [1 0 0] zone axis in Fig. 2a indicates a bcc average lattice. The SAED pattern has high intensities, but no detectable diffuse scatterings are present, indicating a possible homogeneous chemical composition and a fully random bcc solid-solution of the HEA-4E in the as-cast state. The microstructure and SAED for the HEA-5E also shows similar features with no detectable diffuse scattering or any contrast due to a second-phase formation [24]. Fig. 2b is the enlarged Fourier filtered lattice image of the local structure of HEA-4E viewed along the [1 0 0] zone axis. The lattice fringes show some local waviness and distortions similar to that observed for other HEAs like MoNbTaW and CoCrFeNiAl [6, 26], indicating a high level of local lattice distortion. Figs. 2c and 2e are the two boxed regions of Fig. 2b, little enlarged for a detailed analysis of the lattice distortions. Figs. 2d and 2f are the same as Figs. 2c and 2e, but here the lattice fringes are traced with the white lines. It can be seen that the lines tracing the lattice fringes become discontinuous as it approaches regions of severe lattice distortions. At those parts, the lattice fringes could not be traced with continuous lines.

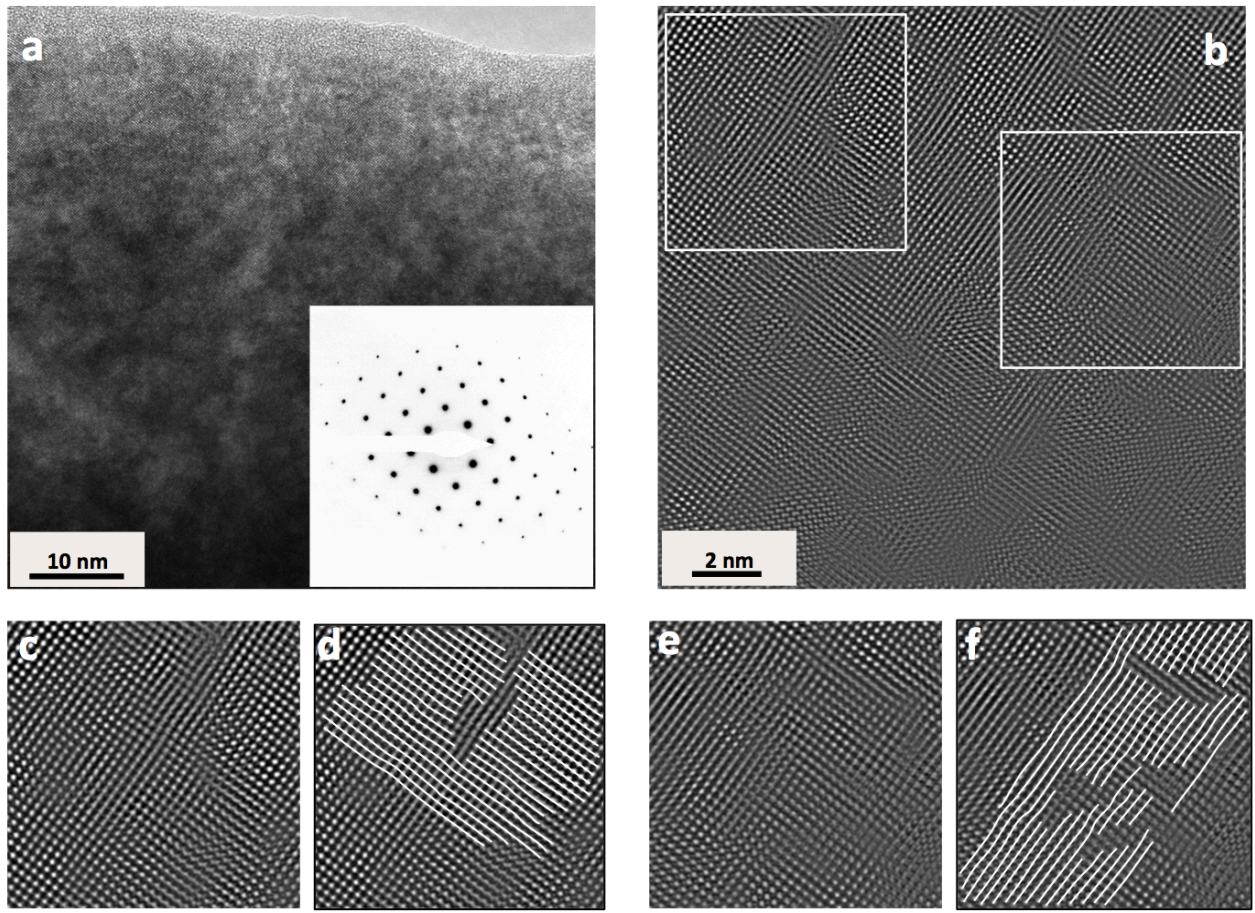

Fig. 2. HRTEM images of the local structure of as-cast TaNbHfZr: (a) bright-field TEM image oriented in [1 0 0] zone axis and the corresponding SAED pattern in inset; (b) lattice image obtained by inverse Fourier transform; (c) and (e) are enlarged boxed regions of (b) showing local lattice distortions; (d) and (f) indicate the local lattice distortions with discontinuous traced lines of lattice fringes marked in white.

The SAEDs of HEA-4E and HEA-5E in their as-cast state apparently indicate an absence of local chemical order, but there still might be some undetected weak diffuse scattering around the Bragg spots due to the SRC in a substitutional solid-solution structure. To analyze the local chemical composition and any traces of SRC in the as-cast state, APT analysis was done on the as-cast HEA-4E. Fig. 3 shows the reconstructed atom-map of the constituting elements of HEA-4E. The atom maps for different elements show a rather homogeneous distribution at the local nanometer scale (Figs. 3a-3d). Fig. 3e shows a cylindrical volume embedded along the long axis of the reconstructed atom-map. This volume is used to create a 1-dimensional (1-D) in-depth concentration profile of the elements as shown in Fig. 3f. The compositions of the different elements almost remain flat throughout the length of the analyzed cylindrical volume. This homogeneity in the local compositions and the lack of observed chemical clusterings from the isoconcentration surfaces in the APT reconstructions indicate an absence of local chemical ordering as also evidenced from the SAED patterns of as-cast HEA-4E (Fig. 2) and HEA-5E [24].

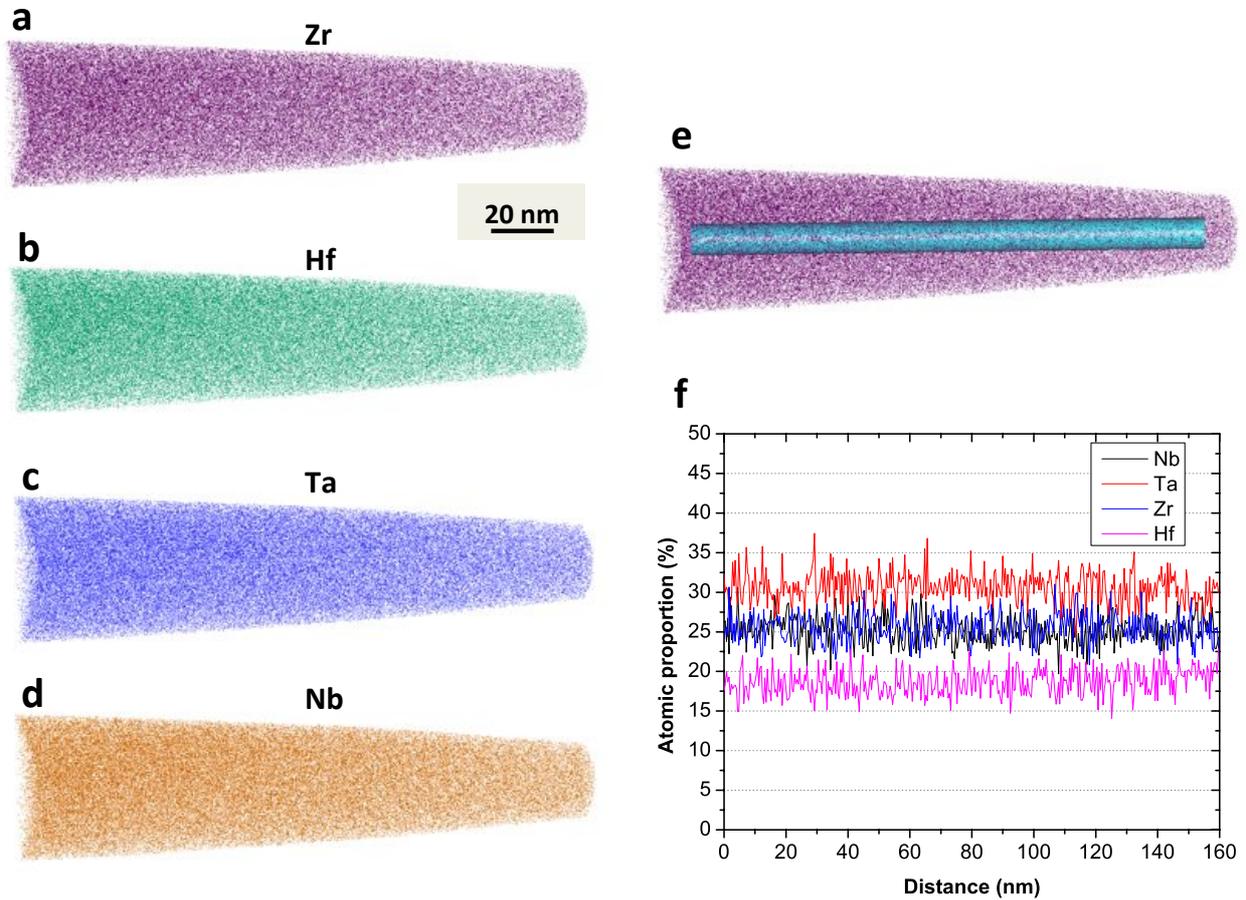

Fig. 3. Reconstructed atom maps and local compositions from the APT run of as-cast TaNbHfZr: (a), (b), (c) and (d) are the atom maps for Zr, Hf, Ta and Nb, respectively; (e) position of the embedded cylinder for 1-D concentration profile; (f) local 1-D composition profile along the axis of the cylinder.

*3.2 Average structure and bulk compositions of the annealed state*

If the as-cast HEA-4E and HEA-5E are annealed at 1800 °C, then the thermodynamic stability parameter $\Omega$ is calculated to be 2.35 and 2.58, respectively, which suggests a stable solid solution structure for the HEAs. These values of $\Omega$ are calculated based on the assumption that $\Delta H_{mix}$ and $\Delta S_{mix}$ remain constant for a random solid solution. This is verified by the powder XRDs done on the two studied HEAs annealed for 1 day. Fig. 4 shows the powder XRD patterns of the annealed HEA-4E and HEA-5E with all the major peaks indexed with a bcc average structure. Little traces of peaks from a hexagonal close packed (hcp) structure are detected, but there is no such major phase decomposition observed. Fig. 5 shows the microstructure and the homogeneity of composition as obtained from the SEM-EDX analysis.

Microstructures taken in the BSE mode offer atomic number (Z) contrast, which helps in identifying traces of local inhomogeneity [27].

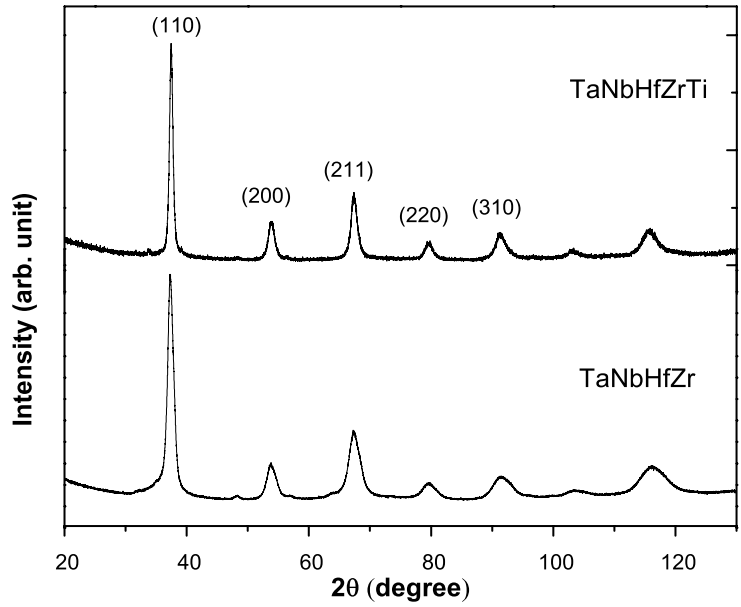

Fig. 4. Powder XRD pattern of the TaNbHfZr and TaNbHfZrTi HEAs annealed for 1 day.

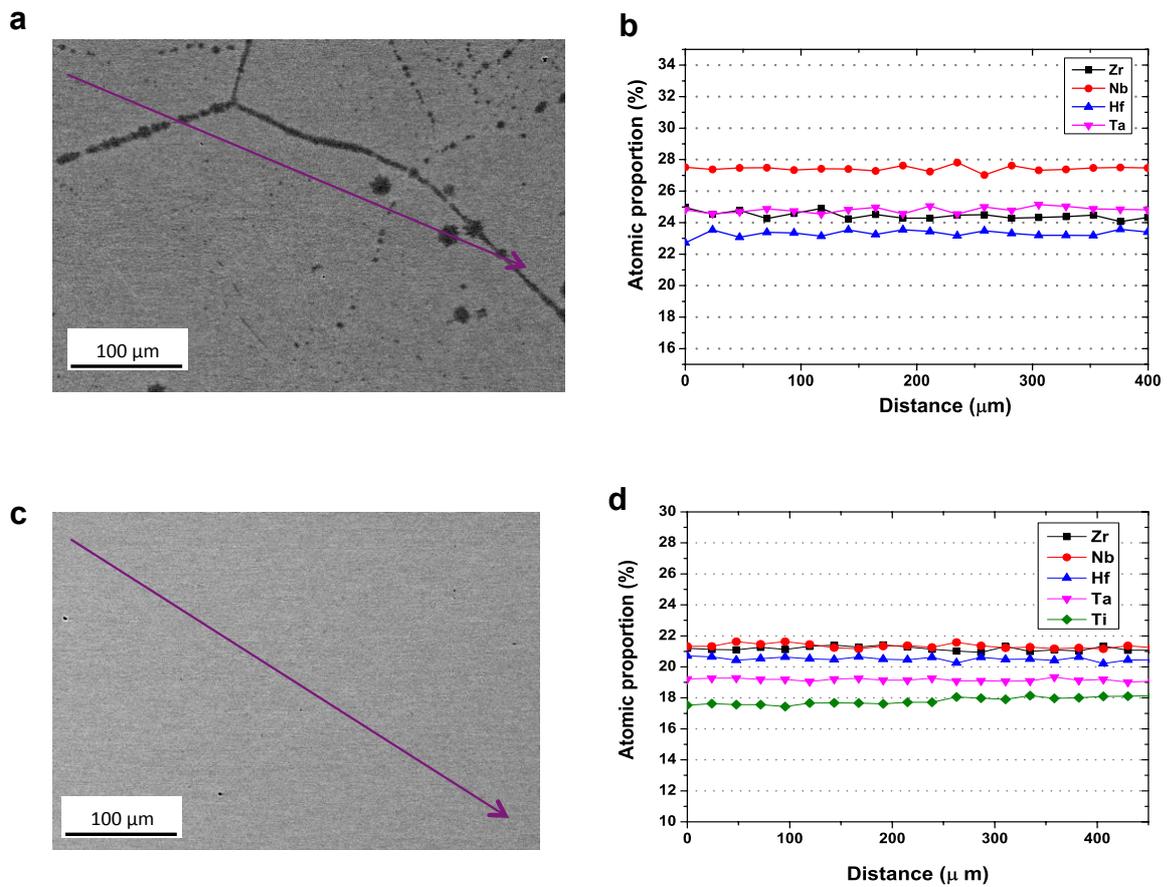

Fig. 5. Microstructure and composition analysis of TaNbHfZr and TaNbHfZrTi annealed for 1 day: (a) and (c) are SEM-BSE images of TaNbHfZr and TaNbHfZrTi, respectively; (b) and (d) are the EDX line scan along the arrows in (a) and (c).

The compositions from the EDX line-scan analysis along the arrows drawn in the BSE images show that the atomic compositions remain almost constant over a range of 400 μm. The elemental compositions vary within a maximum range of 1 atomic %. The annealed HEA-4E shows some traces of the primary and secondary dendrites from the as-cast microstructure [21], whereas the annealed HEA-5E shows almost no Z contrast and traces of inhomogeneity.

### *3.3 Local structure and chemical ordering in the annealed state*

Annealed HEA-4E and HEA-5E produce traces of diffuse scattering in single crystal XRD and SAED, indicating the presence of significant local chemical ordering. The local chemical composition of the annealed HEAs were investigated through the APT techniques as for the as-cast material described before. Fig. 6 represents the elemental atom-map from the APT reconstructions of the HEAs.

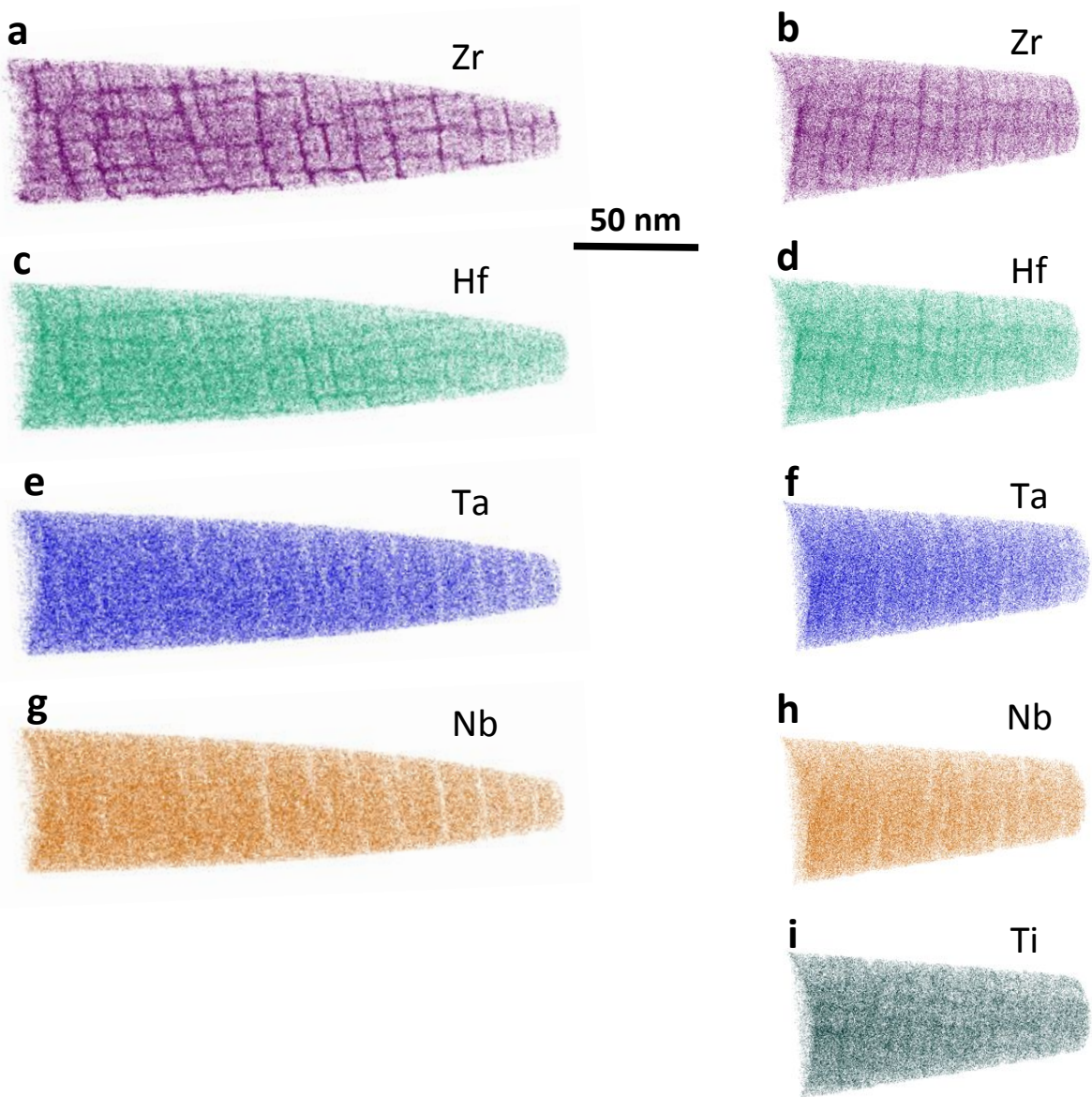

Fig. 6. Atom maps from APT reconstructions of TaNbHfZr and TaNbHfZrTi annealed at 1800°C for 1 day: (a), (c), (e) and (g) are the atoms maps of Zr, Hf, Ta and Nb atoms in TaNbHfZr; (b), (d), (f), (h) and (i) are the atoms maps of Zr, Hf, Ta, Nb and Ti atoms in TaNbHfZrTi.

There appears a 3-dimensional grid-like network of perpendicular filaments of planar chemical clusters of Zr atoms in both the annealed HEA-4E and HEA-5E. A weak co-clustering of Hf atoms along with the Zr clusters is indicated by the atom-maps of both the annealed HEAs. The regions locally enriched with Zr and Hf are also the regions depleted in Ta and Nb, as logically, observed for both the HEAs. The atom map of Ti (Fig. 6i) shows that Ti is rather evenly distributed in HEA-5E, but a slight co-clustering with Zr is evident at some places.

The chemical SRCs detected by the APT are expected to affect the local lattice structure of the HEAs because of the larger atomic volumes of Zr (atomic volume of Ta, Nb, Hf, Zr and Ti being 17.99, 17.98, 22.35, 23.28 and 17.67 $Å^3$, respectively). The local structure is revealed by HRTEM lattice images (Fig. 7). The contrast in the BF lattice image captured along the [1 0 0] zone axis shows local mutually perpendicular regions similar to the contrast from SRCs of APT investigations.

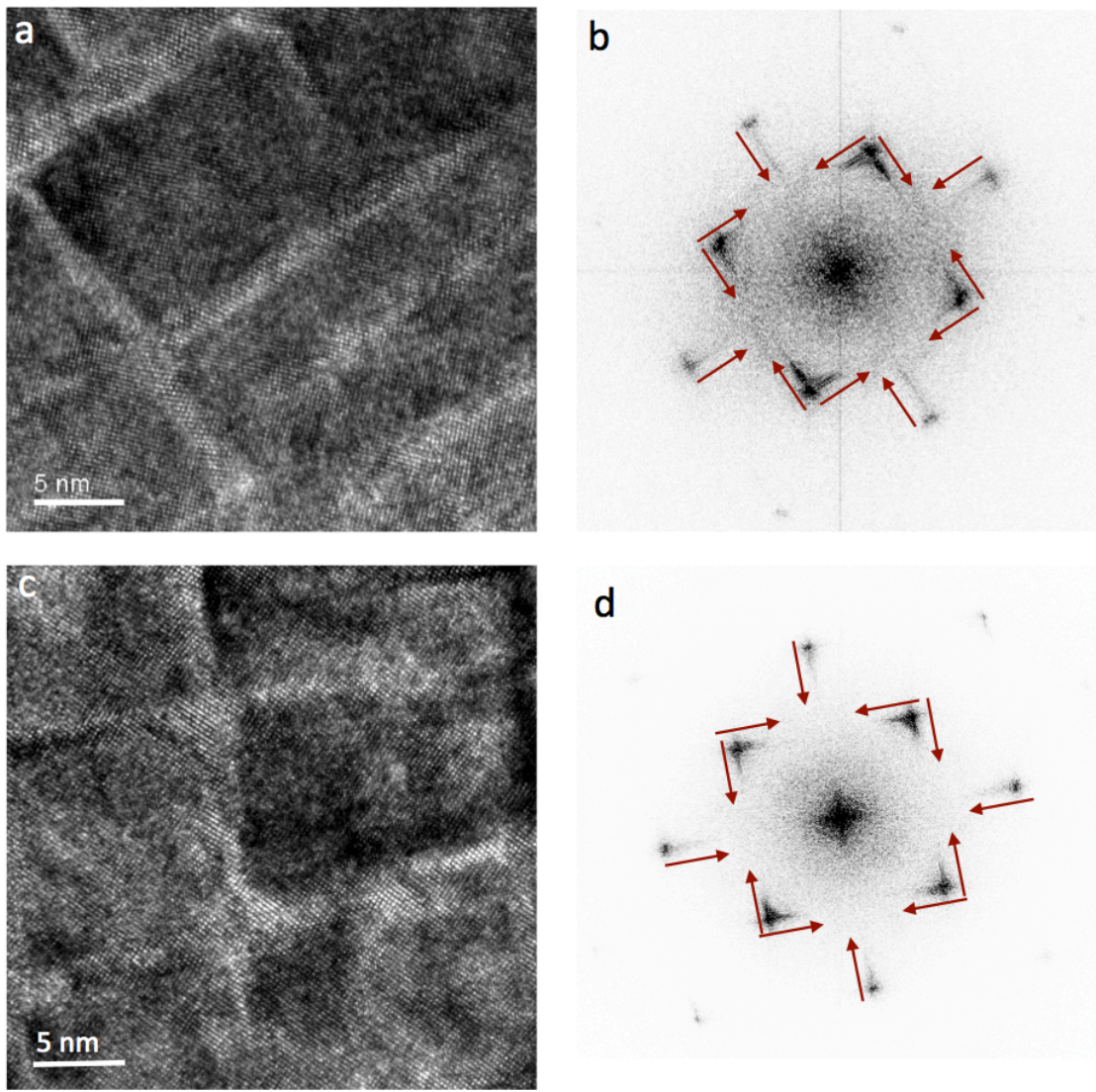

Fig. 7. HRTEM investigations of annealed TaNbHfZr and TaNbHfZrTi: (a) and (c) are the HRTEM lattice images in BF mode of HEA-4E and HEA-5E containing the SRCs; (b) and (d) are the fast Fourier transform (FFT) of (a) and (c). The streak-like features emerging from the Bragg reflection positions are indicated with small arrows.

The lattice-fringe lines passing across the local structure shows a gradual bending at places which have the similar shapes as the SRCs found in APT analysis. This local bending/ relaxation of lattice is caused by the Zr and Hf atoms in the SRCs with larger atomic radius than the average structure appearing perpendicular to the <1 0 0> axes. The streaks emerging asymmetrically towards lower reciprocal vectors in the FFTs are due to the directional local tetragonal lattice relaxations. A detailed study on the relationship of the asymmetric streak-like diffuse intensity and the local lattice relaxations for the TaNbHfZr system is published elsewhere [21].

To understand the general local composition variations of the SRCs evidenced from the atom-maps and the HRTEMs, the chemical clusters have been analyzed by the proximity histogram (proxigram) analysis, which calculates the local compositions of regions at a fixed distance from the surface of the selected cluster. A total of around 11 nm of distance is analyzed starting far away from the selected SRC to the core of the SRC. The selected SRCs and their proxigrams of the two HEAs are shown in Fig. 8. As expected from the atom maps (Fig. 6), Zr atoms cluster in high proportion inside the SRCs.

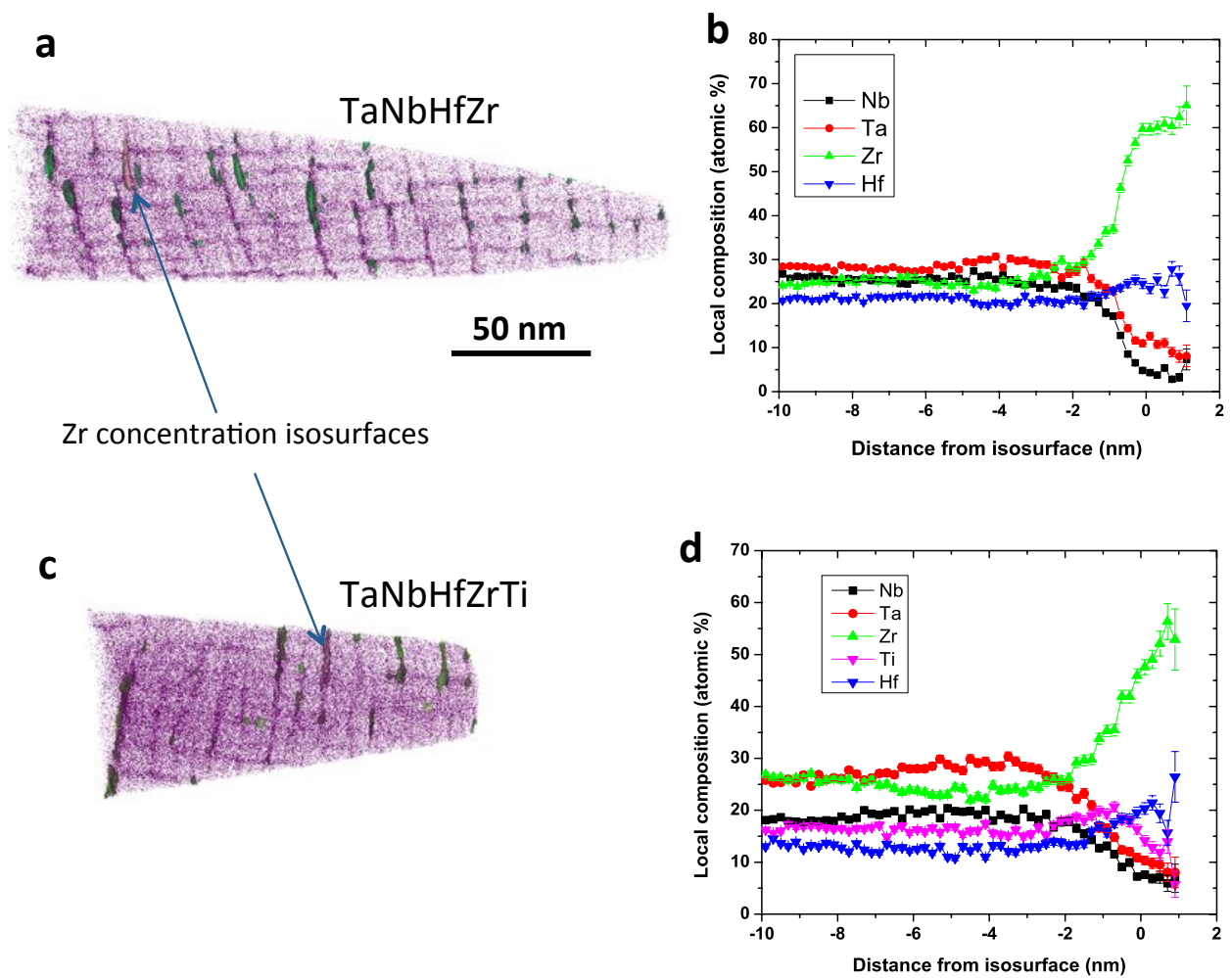

Fig. 8. Local chemical composition analysis of SRCs by proxigram method: (a) and (c) shows the Zr atom map with arrows showing the chemical cluster of TaNbHfZr and TaNbHfZrTi, respectively, which is studied for local compositions; (b) and (d) are the proxigrams based on the Zr concentration isosurfaces shown in (a) and (c).

According to the proxigram analysis, the Zr concentration goes up to around 65 and 55 atomic % for the HEA-4E and HEA-5E, respectively. This is only an approximate value due to many instrumental parameters affecting the running conditions of the APT, but the peak value of the Zr concentration is 160 % and 120 % higher than the composition in the average bulk of the HEA-4E and HEA-5E, respectively. For both the HEAs, the local composition becomes gradually depleted in Ta and Nb as analyzed from regions away from the SRCs to the interior of the SRCs. The Hf concentrations remain almost constant for both the studied HEAs, but it shows only a 5-10 atomic % increase inside the SRCs, showing some tendency of co-clustering of Hf with the Zr atoms. In the HEA-5E, the Ti atoms are distributed almost equally in the proxigram as can also be expected from the atom map (Fig. 6i). At the vicinity of the SRC the Ti concentration goes up by around 5 % only and then drops at the core of the clusters.

All this local chemical information is particularly helpful to construct a real local structure model for thermodynamic, structural and mechanical properties of HEAs with SRO/SRC present in it. Figs. 6-8 show the interdependence of the chemical clustering, local lattice relaxations, lattice-fringe bending and its effect on the streak-like features produced in the FFT of the HRTEM images. Since Fig. 7 just shows some local structural analysis with the related diffuse scattering from the FFTs, it can be intriguing to understand and compare with the average effect of the local structural ordering from the bulk of the HEAs. Fig. 9 shows the structural model built for the HEA-4E, the diffuse scattering in XRD from the bulk sample and the simulated diffuse scattering based on the relaxation of the structural model.

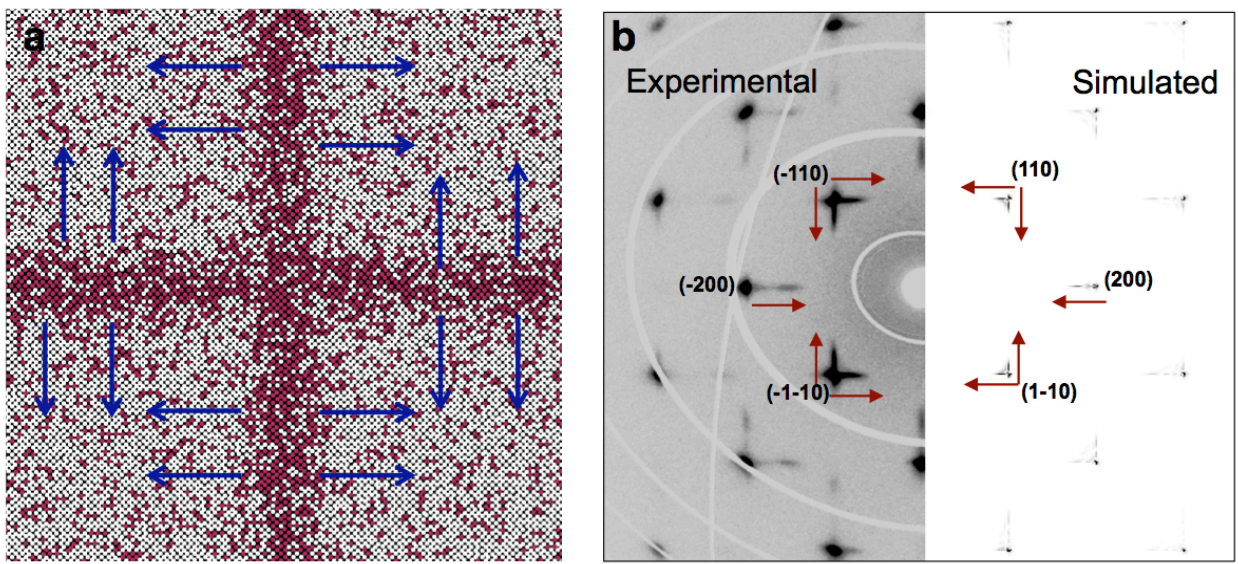

Fig. 9. Atomic model of HEA-4E, its experimental diffuse scattering from the bulk-average local ordering and simulated diffuse scattering: (a) atomic model containing SRC for the MD relaxation with blue arrows showing the direction of tetragonal relaxation, Zr atoms are shown in maroon, (b) experimental diffuse scattering in ($hk0$) reciprocal lattice plane obtained from the single crystal XRD at the synchrotron source; simulated XRD pattern showing streak-like diffuse scattering intensities. The asymmetric extensions of the diffuse scatterings stretching from the main Bragg reflections are shown with maroon arrows.

The local structural model (Fig. 9a) with chemical ordering was constructed from the local composition analysis of proxigrams (Fig. 8b), in-depth compositional analysis across the SRCs and the shape and size of the chemical clusters from HRTEM analysis. The structural model presented here is a supercell of 80×80×20 bcc unit cells containing 256,000 atoms. The local concentration profiles across the SRCs for both HEA-4E and HEA-5E are found to match a Gaussian distribution with FWHM of 2-2.5 nm. In Fig. 9b the experimental diffuse scattering is obtained for the (*hk0*) reciprocal lattice layer reconstructed from a single-crystal XRD dataset collected at the synchrotron source. The simulated reciprocal lattice layer (Fig. 9b) is calculated from the atomic model of HEA-4E (Fig. 9a) relaxed by the embedded atom method (EAM) potential based molecular dynamics (MD) [28]. The details of the experimental procedure related to XRD at synchrotron, the MD relaxation and the calculated reciprocal lattice plane is discussed in our earlier publication [21].

### *3.4 Lattice distortion and strengthening*

In this study, the hardness evolution of HEA-4E and HEA-5E for up to 1 day of annealing is presented in Fig. 10. The grain sizes of the materials were more than 100 μm and any strengthening from the grain size effect can be neglected [29]. The hardness of the HEA-4E and HEA-5E in the as-cast state are found to be 2.94 and 3.40 times that of the expected value from the rule of mixture. This is consistent with the multiple times strengthening observed in fully random solid-solution HEAs without any SRO/SRC in the as-cast and annealed conditions [6, 18-20].

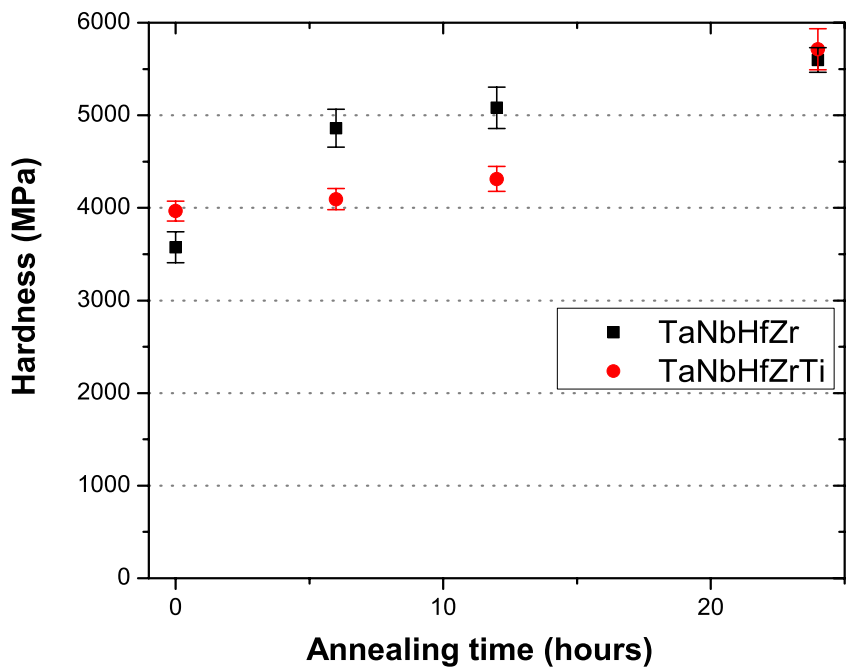

Fig. 10. Change in hardness with annealing time and increasing amount of SRC in HEA-4E and HEA-5E.

This high hardness in the as-cast condition is supposed to be originating only from the solid-solution like strengthening, local lattice distortion or lattice strain effects [4]. The observed local distortions in the lattice fringes of the HRTEM images of HEA-4E in Figs. 2c-2f shows supportive evidence in favor of the mechanical strengthening due to the local lattice distortions. Similar HRTEM features and correlations with the mechanical strengthening have been found for fully random solid-solution HEAs in the literature as well [6, 26]. The average local lattice distortions of the studied HEAs in their as-cast and annealed state are quantified with the ADPs from single crystal XRDs and powder NDs as presented in Table 1. The thermal/ vibrational ADPs are calculated from the rule of mixture of the pure elements. The total refined ADPs given in Table 1 comprises of both the effect of the static and the thermal displacements [30]. It is evident that the static ADPs increase the overall ADPs to be 2-4 times of that expected from only the thermal vibrations. This kind of static atomic displacements causing the local lattice distortions are also reported for other HEAs like MoNbTaW and ZrNbHf in their as-cast and annealed states [6, 31]. In this study the ADPs in the as-cast state are found to be larger than those of the annealed state. This can happen because the SRCs present in the annealed HEAs would deplete the larger Zr atoms in the average structure matrix, which in terms would reduce the local lattice distortions and the static ADP.

Table 1. ADPs of the studied HEAs refined from the single crystal XRD and powder ND

| HEA material | ADP from ND ($Å^2$) | ADP from single crystal XRD ($Å^2$) | Average thermal ADP ($Å^2$) |
|---|---|---|---|
| HEA-4E, as-cast | 0.0257 ± 0.0002 | na | 0.0060 |
| HEA-4E, annealed | 0.0160 ± 0.0005 | 0.0227 ± 0.0016 | ,, |
| HEA-5E, as-cast | 0.0198 ± 0.0005 | na | 0.0059 |
| HEA-5E, annealed | 0.0128 ± 0.0007 | 0.0226 ± 0.0013 | ,, |

* na: not available

In our previous publication it has been shown that the SRCs in HEA-4E locally reduce the lattice potential energy [21]. The local energy minima at the SRCs would provide an additional energy barrier to the dislocation motion against plastic deformation [16]. Based on this reasoning, the calculated strengthening resulting from the SRCs was found to match the maximum value of extra strengthening observed in HEA-4E. The hardness and the compressive yield strength of the HEA-4E increasing with annealing time (until 1 day) was associated with the amount of SRCs formed and their mutual interconnections. The observed increase in hardness in the annealed HEAs is solely due to the increasing amount of SRCs appearing in the single-phase average matrix. Otherwise, the increased grain size and annihilation of quenched-in vacancies in the annealed alloys are only expected to reduce the strength and hardness [29, 32], which is contrary to the observed trend for the studied HEA-4E and HEA-5E. The hardness of the HEA-4E and HEA-5E increases by 57% and 44% after 1 day of annealing, respectively.

# 4. Thermodynamic modeling

The enthalpy of mixing, $\Delta H_{mix}$, used in the thermodynamic parameter, $\Omega$, was calculated by density functional theory (DFT) based calculations to achieve quantum mechanical accuracy [33]. The Vienna *ab inito* simulation package (VASP) was used to calculate the ground state energy of the HEAs by using the generalized gradient approximation (GGA) scheme [34]. Supercells of sizes 5×5×4 and 5×5×5 bcc unit cells were used to calculate the $\Delta H_{mix}$ for the HEA-4E and HEA-5E containing 200 and 250 atoms, respectively. Projector augmented wave (PAW) method [35] with a maximum kinetic energy cutoff of 500 eV for the plane waves was used to calculate the energy of the relaxed structure. The Monkhorst-Pack [36] scheme was used for the Brillouin zone sampling around the Gamma point for the HEA supercells. For calculations of the energies of the pure elements, the Brillouin zone sampling was done with a 12×12×12 grid point sampling for unit cells containing 2 atoms. The energy of the HEA structures was minimized iteratively to a tolerance level of $10^{-5}$ eV/ supercell for electronic steps and $10^{-4}$ eV/ supercell for ionic relaxations, respectively, which amounts to a energy precision level of 0.5 meV/ atom in the calculated structures. The conjugate gradient method was used to minimize the energy by ionic relaxations.

Two different approaches can be found in the literature for the *ab-initio* calculation of the energy of substitutional alloys; firstly, the special quasirandom structures (SQS) [37] approach comprising of a few atoms (16-32) and secondly, the standard supercell structure approach with few hundreds of atoms calculated by modern high-speed computers [38]. SQS was developed mainly to mimic the SRO parameters of binary alloys for a limited number of atoms. For multicomponent alloys the use of the SQS technique can be challenging as the possible configuration space grows with the concentration and the number of elements present in the alloys [38]. Whereas supercells consisting of 108 or 128 atoms have been shown to perform as good as the alloy structures generated by SQS with respect to the internal energy and elastic constants [38]. In the literature, HEA supercells containing 200 atoms were already used for DFT-based ab-intio molecular dynamics and pair-distribution function analysis [39]. The thermodynamic and atomic size parameters, which determine the phase stability in HEAs, are presented in Table 2 for the studied HEA-4E and HEA-5E for their as-cast state. The supercell structures were created by generating random numbers and the configurational entropy is calculated for the fully random solid-solutions.

Table 2. Calculated thermodynamic parameter (Ω), atomic size parameter (δ) and related variables for the phase stability of HEA-4E and HEA-5E in the as-cast state.

| Material | $\Delta H_{mix}$ (meV/atom) | $T_m$ (K) | $\Delta S_{mix}$ ($k_B$) * | δ (%) | Ω (%) |
|---|---|---|---|---|---|
| TaNbHfZr | 105 | 2669 | 1.386 | 5.38 | 3.03 |
| TaNbHfZrTi | 111 | 2523 | 1.609 | 4.92 | 3.14 |

* $k_B$ : Boltzmann constant

Unlike the as-cast state, in the annealed state the HEA-4E and HEA-5E shows marked SRCs with local composition fluctuations (Fig. 8). This kind of local compositional fluctuations are expected to cause extra elastic interactions between the different local configurations and their chemical energies affecting the free energy of the system [40]. Also the change of enthalpy of the average structure would be accompanied by a lowering of configurational entropy [17]. In this work the enthalpy of the alloy is calculated from the potential energy obtained from the MD relaxation of the modeled structure of HEA-4D. The change in enthalpy due to the appearance of SRCs, $\Delta H_{SRC}$ is calculated by

$$\Delta H_{SRC} = E_{SRC} - E_{Rand},$$

where $E_{SRC}$ and $E_{Rand}$ are the potential energies of the relaxed structures containing the SRC and the fully random solid-solution, respectively. The value of $\Delta H_{SRC}$ comes out to be -19.2 meV/ atom averaged over the whole modeled structure, which shows a lowering of the enthalpy in the structure containing the SRCs.

The lowering of configurational entropy due to the presence of SRCs can be determined from the SRO parameters as suggested by Guggenheim [17]. The calculations involve counting bonds of the whole structure for the next neighbor elements and the calculations can become cumbersome for large systems with unknown SRO parameters. Also the method of calculating entropy only considers next-neighbor correlations and may not reveal the true global picture of the system. On the other hand, de Fontaine [40] has shown that the entropic contribution to the free energy in a non-homogeneous system can be determined form the local compositions of the different microstates/ sublattice of the superstructure/ supercell. The local single-site microstates are considered as point-clusters and the generalized Bragg-Williams analysis is applied to derive the average configurational entropy of the whole structure. This kind of analysis based on the single-site compositions have been also found to be successful in deriving the thermodynamic energy functional description of non-homogeneous solutions as established by Cahn and Hilliard [41].

The local single-site configurational entropy average of the HEA-4E structure calculated for the structure model used for MD simulations comes out to be 1.3035 $k_B$/ atom. This is lower than the maximum value of configurational entropy $ln(4) = 1.386$ $k_B$/ atom in the fully random structure. The change in Gibbs free energy, $\Delta G$ can be expressed as

$$\Delta G = \Delta H - T\Delta S,$$

where $\Delta H$ is the change in enthalpy, $\Delta S$ is the change in entropy and $T$ is the temperature of the system. Applying this formulism between the as-cast and annealed state of HEAs, the $\Delta H$ becomes $\Delta H_{SRC}$, as evolution of SRC is the only structural change observed. Similarly, the $\Delta S$ can be assumed as $\Delta S_{conf}$, the change in configurational entropy due to the occurrence of chemical ordering, considering other sources of entropy like the vibrational and electronic entropy remain unchanged. So the expression for $\Delta G$ becomes

$$\Delta G = \Delta H_{SRC} - T\Delta S_{conf},$$

where $\Delta H_{SRC}$, the change in enthalpy contribution is calculated to be be -19.2 meV/ atom, $\Delta S_{conf}$ becomes (1.3035 - 1.386) = -0.0825 $k_B$/ atom, $T$ is the temperature of annealing 1800°C or 2073 K. This gives the change in entropic contribution to the Gibbs free energy, $-T\Delta S_{conf}$ as 14.7 meV/ atom. So, the total change in Gibbs free energy, $\Delta G$ due to enthalpy and entropic contribution becomes (-19.2 + 14.7) = -4.5 meV/ atom, favoring the formation of the observed SRCs in the annealed HEAs.

## Conclusions

HEAs like TaNbHfZr and TaNbHfZrTi form a bcc solid-solution in the as-cast condition and after annealing at 1800°C for one day. The bcc structures should be stable in both the as-cast and annealed states according to the calculated thermodynamic and atomic size related factors relevant for the HEAs. However, the APT analysis and the diffuse scattering in the single-crystal XRDs reveal the presence of SRC type chemical order after annealing. The SRCs are most enriched in Zr, which have a Gaussian-like concentration profile across the width of the planar clusters perpendicular to the <1 0 0> axes. The SRCs distort the lattice by local tetragonal relaxations producing streak-like diffuse scattering intensities adjacent to the Bragg reflections in the single-crystal XRDs and FFTs of HRTEM lattice images. The as-cast HEAs are random solid-solutions without any detectable local order. But, the lattice in the as-cast alloys is locally distorted as observed through the frequent waviness and discontinuities of the lattice fringes caused by the high degree of static atomic displacements. The observed hardness in as-cast condition is multiple times higher due to the local lattice distortion related solid-solution hardening. With the increase in annealing time, the hardness further increases by around 50%. The increasing amount of SRCs appearing with longer annealing times act as local energy barrier for the dislocation motion. The extra shear stress needed for the dislocations to overcome the local lattice potential energy barriers increase the hardness of the bulk material. The presence of SRC reduces the enthalpy of the relaxed structure compared to the random solid-solution. Thermodynamic modeling shows that the effect of the reduction of enthalpy due to SRC can balance the reduced entropic contribution to the Gibbs free energy. This makes the annealed structures to contain SRCs and still reduce the free energy of the system.

**Acknowledgements**

The authors would like to thank Dr. S. Gerstl and Dr. R. Schäublin (SCOPEM, ETH Zurich) for the technical help in APT and HRTEM; Dr. V. Pomjakushin (SINQ, PSI, Villigen Switzerland) for the support in the powder neutron diffraction. Financial support of Swiss National Science Foundation under the grant 200020_144430 is gratefully acknowledged.